\documentclass[]{aa} 
\usepackage{graphicx}
\usepackage{txfonts}
\begin{document}
%
\def\hi {H\,{\sc i}}
\def\hdueo {H$_2$O}
\def\meth {CH$_{3}$OH}
\def\dg{$^{\circ}$}
\def\kms{km\,s$^{-1}$}
\def\jyb{Jy\,beam$^{-1}$}
\def\mjyb{mJy\,beam$^{-1}$}
\def\solmass {\hbox{M$_{\odot}$}}
\def\solum {\hbox{L$_{\odot}$}} 
\def\d {$^{\circ}$}
\title{Methanol masers probing the ordered magnetic field of W75N}

\author{G.\ Surcis \fnmsep\thanks{Member of the International Max Planck Research School (IMPRS) for Astronomy and Astrophysics at the Universities of Bonn and Cologne}\,  \inst{1,2}
  \and 
   W.H.T. \ Vlemmings \inst{1}
\and 
   R. \ Dodson \inst{3}
\and
   H.J. \ van Langevelde \inst{4,5}}
\offprints{G. Surcis}

\institute{ Argelander-Institut f\"{u}r Astronomie der Universit\"{a}t Bonn, Auf dem H\"{u}gel 71, 53121 Bonn, Germany\\
 \email{gsurcis@astro.uni-bonn.de}
 \and
 Max-Planck Institut f\"{u}r Radioastronomie, Auf dem H\"{u}gel 69, 53121 Bonn, Germany
 \and
 International Centre for Radio Astronomy Research, University of Western Australia, Perth, Australia
 \and
 Joint Institute for VLBI in Europe, Postbus 2, 79990 AA Dwingeloo, The Netherlands
 \and
 Sterrewacht Leiden, Leiden University, Postbus 9513, 2300 RA Leiden, The Netherlands}

\date{Received ; accepted}
\abstract
{The role of magnetic fields during the protostellar phase of high-mass star-formation is a debated topic. In particular, it is still unclear how magnetic fields influence the formation and dynamic of disks and outflows. Most current information on magnetic fields close to high-mass protostars comes from \hdueo \, and OH maser observations. Recently, the first 6.7\,GHz methanol maser polarization observations were made, and they reveal strong and ordered magnetic fields.}
{The morphology of the magnetic field during high-mass star-formation needs to be investigated on small scales, which can only be done using very long baseline interferometry observations. The massive star-forming region W75N contains three radio sources and associated masers, while a large-scale molecular bipolar outflow is also present. Polarization observations of the 6.7\,GHz methanol masers at high angular resolution probe the strength and structure of the magnetic field and determine its relation to the outflow.} 
{Eight of the European VLBI network antennas were used to measure the linear polarization and Zeeman-splitting of the 6.7\,GHz methanol masers in the star-forming region W75N.}
{We detected 10 methanol maser features, 4 of which were undetected in previous work. All arise near the source VLA\,1 of W75N. The linear polarization of the masers reveals a tightly ordered magnetic field over more than 2000 AU around VLA\,1 that is exactly aligned with the large-scale molecular outflow. This is consistent with the twisted magnetic field model proposed for explaining dust polarization observations. The Zeeman-splitting measured on 3 of the maser features indicates a dynamically important magnetic field in the maser region of the order of 50\,mG. We suggest VLA\,1 is the powering sources of the bipolar outflow.}
{}

\titlerunning{W75N: methanol masers polarization.}
\authorrunning{Surcis et al.}

\maketitle
\section{Introduction}
Magnetic fields are attributed an important role in star formation, in particular in halting the collapse, transferring angular momentum, and powering the outflows. Gravitational collapse of molecular clouds proceeds primarily along the magnetic field lines, giving rise to large rotating disk or torus structures orthogonal to the magnetic field (Matsumoto \& Tomisaka \cite{mat04}). Consequently, the molecular bipolar outflows, which originate in the protostar, are driven parallel to the magnetic field (e.g., Cohen \cite{coh05}; McKee \& Ostriker \cite{McK07}). Despite it is largely accepted for low-mass star-formation, this picture is still under debate for the formation of high-mass stars.\\ 
\indent The best probes of the magnetic field on the smallest scales in the high-mass star-forming environment are maser lines. Their bright and narrow spectral line emission is ideal for detecting both the Zeeman-splitting and the direction of the magnetic field. So far, the most investigated sources are \hdueo \, and OH masers, revealing an ordered structure in the magnetic field and field strengths between 10 and 600 mG (e.g., Vlemmings et al. \cite{vle06a}) and a few mG (e.g., Fish et al. \cite{fish05}), respectively. Even though methanol (\meth) is the most abundant of the massive star-formation maser species (e.g. Pestalozzi \cite{pes07}), the first 6.7 GHz ($5_{1}\rightarrow6_{0}A^{+}$), methanol maser linear polarization observations have only recently been made (e.g., Vlemmings \cite{vle06b}; Green et al. \cite{gre07}; Dodson \cite{dod08}). Like \hdueo, \meth \, is a non-paramagnetic molecule, and thus any linear polarization is a few percent ($\sim$2--3\%), while circular polarization is well below 1\%. Still, Vlemmings (\cite{vle08}) detected 6.7 GHz methanol maser Zeeman-splitting in a sample of 24 massive star-forming regions, including W75N, for the first time using the Effelsberg telescope. This revealed a magnetic field strength in the methanol region of about \,20\,mG.\\

\indent In Fig. 1 we show the area of interest of this letter. VLA\,1 and VLA\,2, which were studied in detail by Torrelles et al. (\cite{tor97}), are part of the region W75N(B) in the active high-mass star-forming region W75N (Hunter et al. \cite{hun94}) located at a distance of 2~kpc. A large-scale, high-velocity outflow, with an extension greater than 3 pc and a total molecular mass greater than 255\,\solmass,  was detected and mapped at different frequencies (e.g., Shepherd et al. \cite{she03} and references therein). Shepherd et al. (\cite{she03}) propose a multi-outflows scenario where VLA\,2 may drive the large-scale outflow, and VLA\,1 is the centre of another small flow. However, so far, it has been impossible to determine which is the main powering source of the 3 pc outflow.\\
\indent Several maser species were detected in W75N(B): \hdueo \, (e.g., Torrelles et al. \cite{tor97}), OH (e.g., Baart et al \cite{baa86}, Slysh et al. \cite{sly02}), and \meth \, (Minier et al. \cite{min00}, hereafter MBC). The methanol masers were detected in a linear structure and off-set from the compact clump at northwest (group A) and south (group B) of VLA\,1, respectively (MBC). VLA\,2 is the place where the most intensive OH flare took place (Alakov et al. \cite{ala05}). Other OH maser emission sites are situated on a ring structure around the radio sources (Hutawarakorn et al. \cite{hut02}).\\
\indent To investigate the magnetic field of this region, a number of OH maser polarization observations have been made (e.g., Hutawarakorn et al. \cite{hut02}; Slysh et al. \cite{sly02}; Fish \& Reid \cite{fis07}). The magnetic field strength obtained from the OH maser polarization was 7\,mG (e.g., Hutawarakorn et al. \cite{hut02}; Slysh et al. \cite{sly02}). During the OH maser flare near VLA\,2, a strong magnetic field of more than 40 mG was detected in several maser spots (Slysh \& Migenes \cite{sly06}). Effelsberg observations of the 6.7\,GHz methanol masers indicate a magnetic field $B_{\|}=5.7\,\rm{mG}$ and $B_{\|}=9.5\,\rm{mG}$ in the A and B groups, respectively (Vlemmings \cite{vle08}).\\
\begin {table*}[t!]
\caption []{Results.}
\begin{center}
\scriptsize
\begin{tabular}{c c c c c c c c c c}
\hline
\hline
\\
Features & RA      & Dec     & Peak Flux   & $S(\nu)$ &$V\rm{_{lsr}}$ & $P\rm{_{l}}$ & $\chi$       & $\Delta V_{\rm{Z}}$ & $B_{||}$ \\
         &               &               & Density (I) &                    &               &              &              &                          &     \\
         & (J2000)       & (J2000)       & (\jyb)      & (Jy)      &(\kms)         & (\%)         & ($^{\circ}$) &  (m s$^{-1}$)             & (mG)\\
\\
\hline
\\

A1       & 20h38m36.412s & 42d37m35.285s & 15.54       & 20       & 3.54          & $2.0\pm0.4$ & $-22\pm1$  &  -              & - \\
A2       & 20h38m36.409s & 42d37m35.283s & 48.25       & 72       & 4.07          & $4.0\pm1.2$ & $-20\pm8$  &  -              & - \\
A3       & 20h38m36.405s & 42d37m35.236s & 67.24       & 84       & 4.60          & $4.5\pm0.3$ & $-9.6\pm0.3$ &  -            & - \\
A4       & 20h38m36.400s & 42d37m35.165s & 47.58       & 62       & 5.82          & $0.9\pm0.3$ & $-6\pm6$   &  $0.80\pm0.03$  & $16.3\pm0.6$\\
A5       & 20h38m36.401s & 42d37m35.154s & 39.39       & 51       & 5.12          & $3.5\pm1.5$ & $-2\pm12$  &  $0.75\pm0.13$  & $15.2\pm2.6$ \\
B1       & 20h38m36.417s & 42d37m34.846s & 95.38       & 107      & 7.23          & $1.5\pm0.5$ & $-18\pm9$  &  $0.53\pm0.04$  & $10.9\pm0.8$\\
C1       & 20h38m36.418s & 42d37m34.693s & 2.33        & 3        & 9.34          &  -          &      -     &  -              & - \\
C2       & 20h38m36.420s & 42d37m34.545s & 7.02        & 8        & 6.88          &  -          &      -     &  -              & - \\
C3       & 20h38m36.411s & 42d37m34.376s & 4.11        & 5        & 9.16          & $1.3\pm0.3$ & $-25\pm4$  &  -              & - \\
C4       & 20h38m36.403s & 42d37m34.323s & 10.83       & 11       & 9.51          & $1.4\pm0.3$ & $-30\pm21$ &  -              & - \\

\\
\hline
\end{tabular}
\end{center}
\scriptsize{}
\label{masers}
\end{table*}
\indent After investigating the linear polarization of OH masers (1.6 and 1.7\,GHz), Hutawarakorn \& Cohen (\cite{hut96}) suggested that the magnetic field is oriented along the outflow. However, OH masers at 1.6 and 1.7\,GHz are very susceptible to both internal and external Faraday rotation, making a direct determination of the intrinsic magnetic field structure difficult. Hence, the study of the linear polarization of the 6.7 GHz methanol maser can provide more precise polarization angles and, consequently, a more accurate determination of the magnetic field than obtained at lower frequencies. Here we present the first European VLBI network (EVN) 6.7 GHz methanol maser polarization observations.
\section{Observations and data reduction}\label{obssect}
W75N(B) was observed in full polarization spectral mode at 6668.518 MHz with 8 of the EVN antennas on June 14$^{\rm{th}}$ 2008. The bandwidth was 2 MHz, providing a velocity coverage of $\sim100$\,\kms. The data were correlated twice. First, with modest spectral resolution (512 channels; channel spacing $\sim0.2$\,\kms), which enabled us to generate all 4 polarization combinations (RR, LL, RL, LR). The second correlation run with high spectral resolution (1024 channels; channel spacing $\sim0.1$\,\kms) only provided the circular polarization combinations (LL, RR). This high spectral resolution dataset was used to attempt the detection of the Zeeman-splitting with a similar RCP-LCP cross-correlation method that was used for the Effelsberg observations by Vlemmings (\cite{vle08}). The observation time was 8 hours, including overheads on the calibrators 3C48, J2007+4029, 3C454.3, and J2202+4216. The last two were also used for polarization calibration.\\
\indent The data were reduced using the Astronomical Image Processing Software package (AIPS). The initial polarization calibration was performed on J2202+4216. Fringe-fitting and self-calibration were performed on the brightest spectral channel, corresponding to the strongest maser feature ($\sim$\,90 \jyb \, at $V\rm{_{lsr}}=\,7.23$\,\kms). The observations of the polarization calibrators and the unpolarized calibrator J2007+4029 indicate the remaining polarization leakage after calibration is much less than 1\%. Then we imaged the I, Q, and U cubes (2048 $\times$ 2048 mas) of the source W75N(B) using the task IMAGR (beam of 8.75$\times$6.45 mas). The data cubes were combined to produce cubes of polarized intensity and polarization angle. With the exception of the bandpass table, which had to be remade, the calibration solutions were copied from the modest resolution dataset and applied to the high spectral resolution data. From this dataset we produced RR and LL image cubes.\\
\indent Our observations were obtained between two VLA polarization calibration observations\footnote{http://www.aoc.nrao.edu/astro/calib/polar} during which the linear polarization angle of J2202+4216 changed abruptly from $-30^{\circ}$ to $-80^{\circ}$. However, we were able to calibrate the linear polarization angle using 3C454.3, which had a constant polarization angle of 6$^{\circ}\!\!.9$ from VLA monitoring during our observation$^1$.\\
\indent The absolute position of the brightest feature was obtained through fringe rate mapping using the AIPS task FRMAP. We determined its position to be $\alpha_{2000}=20^{\rm{h}}38^{\rm{m}}36^{\rm{s}}\!.417$, $\delta_{2000}=+42^{\circ}37'34''\!\!.846$  with an absolute position error of about 20 mas. The relative position errors depend on the maser flux of each feature, but are better than 1 mas.\\
\begin{figure*}[t!]
\centering
\includegraphics[width = 12 cm, angle = -90]{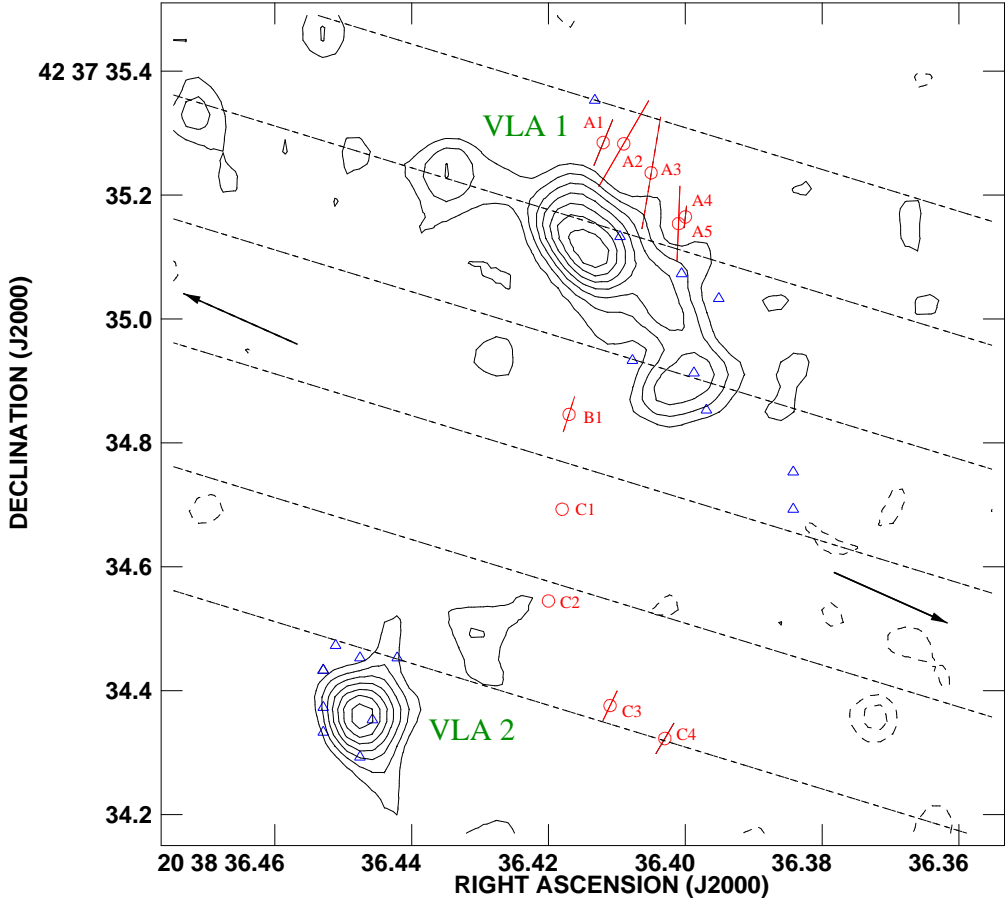}
\caption{Positions of methanol and water masers superimposed on the 1.3 cm continuum contour map of the VLA1 thermal jet and VLA\,2 (Torrelles et al. \cite{tor97}). Contours are $-3$, $-2$, 2, 3, 4, 5, 6, 7, 8 $\times$ 0.16 \mjyb. Blue triangles indicate the positions of the water masers as reported by Torrelles et al. (\cite{tor97}). Red circles indicate the position of the 10 methanol maser features with their linear polarization vectors (40 mas correspond to a linear polarization fraction of 1\%). The two arrows indicate the direction of the bipolar outflow (66\d) and the parallel dashed lines the magnetic field lines (73\d$\pm$\,10\d) as derived from the linear polarization.}
\label{pos}
\end{figure*}
\section{Results}
\begin{figure}[h!]
\centering
\includegraphics[width = 8 cm]{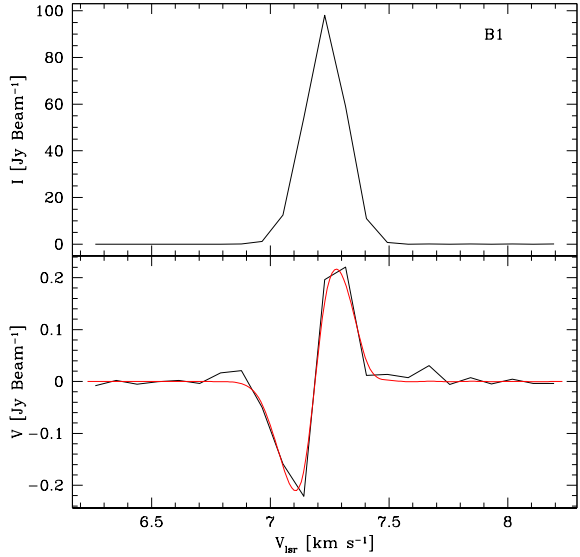}
\caption{Total intensity and circular polarization spectrum for the brightest methanol maser feature of W75N(B). The thick red line in the bottom panel is best-fit total power derivative of a Gaussian fit of I spectrum to the circular polarization spectrum.}
\label{B1}
\end{figure}

\indent We have detected 10 methanol maser features (see Table \ref{masers}), with a minimum signal-to-noise ratio of approximately 80. The peak flux and the flux density ($S(\nu)$) were obtained from a Gaussian fit to the brightest emission channel. The velocity with respect to the local standard of rest ($V_{\rm{lsr}}$) was obtained from a fit to the feature spectrum, as was the linear polarization fraction ($P_l$). The error on the linear polarization angle ($\chi$) includes the formal error due to the thermal noise (Wardle \& Kronberg \cite{war74}). Figure \ref{pos} shows the positions of methanol maser features. They can be divided into three groups: A, B, and C, following the naming convention adopted by MBC. Using three EVN antennas in May 1997, they detected 12 methanol maser features in W75N(B) divided into two groups: A, which is located at the northwest of VLA\,1, and B to the south of VLA\,1, with velocities between 3 and 7 \kms.\\
\indent As found by MBC, the features A1-A5, with velocities between 3 and 6\,\kms, are located along a linear structure. Although the absolute positions in MBC are uncertain by up to 1 arcsec, we were able to identify 5 features detected by MBC as A1-A5. Their velocities are slightly higher ($+0.04$\,\kms$<\Delta\,V\rm{_{lsr}}<+0.16$\,\kms) than those reported previously. Only feature A5 ($V\rm{_{lsr}}$=5.12\,\kms) has a lower velocity ($\sim-0.5$\,\kms). Feature B1, which is located southeast of the group A at a distance of about 400\,mas ($\sim$\,800\,AU at a distance of 2\,kpc), has a velocity of 7.23\,\kms\,. It can be associated with the brightest feature (peak flux\,=\,71.1\,Jy) of MBC. Comparing the separations between A1-A5 and B1 in the two epochs, we find a small difference in their relative positions, implying a separation velocity of $0.93\,\rm{mas\,yr^{-1}}\sim9.5\,$\kms\,. The group C, located south of the feature B1, is composed of four methanol maser features undetected by MBC. The features of this group (C1-C4) seem be along an arc structure with $V_{\rm{lsr}}$ between 6 and 10\,\kms.\\
\indent Linear polarization was detected in 8 of the 10 maser features given in Table \ref{masers}. The highest fractional linear polarization ($P_{\rm{l}}\sim2-5\%$) was detected in group A. The linear polarization fraction of the masers in group B and C is slightly lower ($P_{\rm{l}}\sim1-2\%$). No dependence of $P_{\rm{l}}$ on maser brightness is found. The weighted polarization angle of methanol maser vectors is $\langle\chi\rangle=-17^{\circ}\pm10^{\circ}$. Assuming the interstellar electron density  and the magnetic field are $n_{e}\approx0.012\,\rm{cm^{-3}}$ and $B_{||}\approx2\,\rm{\mu G}$, respectively, the Faraday rotation at 6.7~GHz is $\phi\approx5$\d. Zeeman-splitting was detected in 3 of 10 maser features (A4, A5, and B1). Figure \ref{B1} shows the total intensity and the circular polarization spectrum for the brightest feature. Columns 9 and 10 of Table \ref{masers} contain the Zeeman-splitting in m s$^{-1}$ and the magnetic field strength along the line of sight, which were determined from the cross-correlation between the RR and LL spectra. $B_{||}$ was derived using a Zeeman-splitting coefficient of $0.049$\,\rm{\kms} $\rm{G^{-1}}$ (Vlemmings et al. \cite{vle06b}).
\section{Discussion}
\subsection{\meth~masers distribution}
The 6.7\,\,GHz methanol maser emission likely arises when a methanol-rich cool gas ($T_{\rm{k}}=30-100$\,K) at moderate densities ($10^{3}\,\rm{cm^{-3}}\,<n_{H_{2}}<\,10^{9}\,\rm{cm^{-3}}$) is pumped by the infrared emission from a layer of warm dust (Cragg et al. \cite{cra05}). The warm dust is potentially associated with shocks that originate from the interaction of the VLA\,1 jet and the surrounding medium. These are the shocks that give rise to the W75N(B) \hdueo \, masers that are found in elongated structures near VLA\,1 (Torrelles et al. \cite{tor97}). The velocities of water masers are more extreme by about 8 \kms \, than those of the nearby methanol maser features A1-A5, and the projected distance to VLA\,1 of the \hdueo \, masers is shorter, suggesting that the methanol masers arise in the pre-shock region.\\
\indent As OH and methanol masers often occur in similar environments, the methanol masers could be occurring in a structure related to the large OH maser ring around the three continuum sources (Hutawarakorn et al. \cite{hut02}). The methanol masers display a similar velocity gradient to the OH, with their velocity offset by only $\sim2$\,\kms. However, no methanol emission is detected towards the southern part of the OH ring, making it more likely that the methanol maser structure is related to VLA\,1.
\subsection{The magnetic field}
\begin{figure}[t!]
\centering
\includegraphics[width = 8 cm]{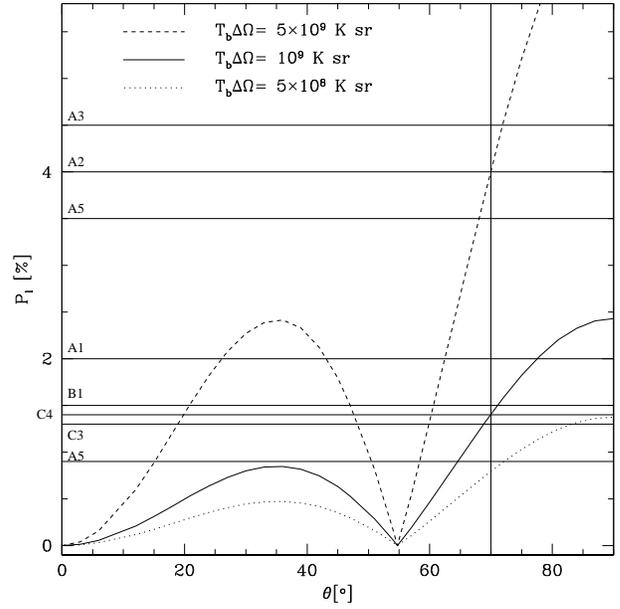}
\caption{The angle $\theta$ between the maser propagation direction and the magnetic field vs. the fractional linear polarization $P_l$ for different values of emerging maser brightness temperatures $T_b\Delta\Omega$. The dashed and dotted curves correspond to the brightness temperature limits for our observations (assuming $\Delta\Omega=10^{-1}$). The horizontal solid curves show the fractional linear polarization of methanol maser features as reported in Table \ref{masers}. The vertical solid line shows the most likely value of $\theta$ for the W75N methanol maser observations.}
\label{theta}
\end{figure}

\indent Before we can relate the observed polarization to the magnetic field, we argue that the polarization vectors are perpendicular to the magnetic field and that the polarization is not affected by non-magnetic phenomena. As in the case for \hdueo \, and SiO masers, the linear polarization of the methanol masers is parallel or perpendicular to the magnetic field, depending on the angle between the magnetic field and the line of sight ($\rm{\theta}$). We solved the polarized radiative transfer using the equations from Nedoluha \& Watson (\cite{ned92}) for the 6.7~GHz methanol maser transition. In Fig.~\ref{theta} we show the relation between $\theta$ and the fractional linear polarization. This relation depends on the emerging maser brightness temperature $T_b\Delta\Omega$. We consider an upper limit of the maser beaming angle $\Delta\Omega\leq10^{-1}$ (Minier et al. \cite{min02}) and the brightness temperature given from 
\begin{equation}
\frac{T_b}{[\rm{K}]}=\frac{S(\nu)}{[\rm{Jy}]}\cdotp \left(\frac{\Sigma^2}{[\rm{mas^2}]}\right)^{-1}\cdotp\xi~,
\end{equation}
where $S(\nu)$ is the flux density, $\Sigma$ the maser angular size, and $\xi$ a constant factor that includes all constant values, such as the Boltzmann constant, the wavelength, and the proportionality factor obtained for a Gaussian shape by Burns et al. (\cite{bur79}),
\begin{equation}
 \xi=13.63~\cdotp10^9 ~\rm{mas^{2} ~Jy^{-1} ~K}.
\end{equation}
Since we find that all maser features are marginally resolved, we take a typical angular size value $\Sigma=7$~mas. Most of the maser features have $T_b\Delta\Omega<10^9$~K~sr, so we find that $\theta<\theta_{\rm crit}\sim55$\d~ is ruled out across the entire methanol maser region. As shown in Goldreich et al. (\cite{gol73}), this means that the linear polarization vectors are perpendicular to the direction of the magnetic field. As indicated in Fig.\ref{theta}, considering the observed brightness temperatures and assuming $\Delta\Omega=10^{-1}$, we estimate the average angle between the magnetic field and the line of sight to be $\theta\sim70$\d. For the non-paramagnetic molecules, however, there are a number of mechanisms that can create linear polarization and/or rotate the polarization angle. The incident infrared photons on the masing region are anisotropically distributed, potentially producing high fractional linear polarization ($P_{\rm{l}}$), oriented tangentially to the incident radial photons. However, since the value of Zeeman frequency shift $g\Omega$ calculated for a magnetic field of 50\,mG is greater than the rate of stimulated emission $R$ (Vlemmings et al. \cite{vle09}), the linear polarization vectors still directly relate to the magnetic field. Consequently, in our case the magnetic field is oriented NE-SW with a position angle of about $73^{\circ}\pm10^{\circ}$. The magnetic field is thus almost perfectly aligned with the large-scale molecular outflow, which is thought to be oriented close to the plane of the sky at a position angle of $66^{\circ}$ (Hunter et al. 
\cite{hun94}). Following our argument that the methanol masers are related to VLA\,1, the alignment of the magnetic field with the large-scale molecular outflows indicates that VLA\,1 is likely the driving source of the outflow.\\
\indent The dust polarization of W75N has been observed at three wavelengths (450, 800, 1100 $\rm{\mu m}$; Greaves \& Holland \cite{gre98}). These observations indicate a change in the magnetic field position angle on the sky from 37\dg ~to ~--35\dg ~with increasing beam size. This change is attributed to the twisted magnetic field lines around the massive protostar, a feature common to the models of low-mass star formation (e.g. Shu et al. 1995). In this case, the large beams are dominated by the dragged-in magnetic field lines, while the highest-resolution observations trace the components of the field that constrain the outflow (Greaves \& Holland \cite{gre98}). Our observations at mas-resolution are consistent with this model and suggest an outflow-driving mechanism similar to the one during low-mass star formation.\\
\indent The magnetic field derived from our EVN observations is between 2 times larger than the one obtained using the Effelsberg telescope (Vlemmings \cite{vle08}). This difference is expected, since the lower spatial resolution observations are often affected by spectral blending (e.g., Sarma et al. \cite{sar01}). Assuming $\theta\sim70^{\circ}$, the actual magnetic field strength in the methanol region is around 50\,mG, which is similar to the region of the OH maser flare (Slysh \& Migenes \cite{sly06}).  Crutcher (\cite{cru99}) found an empirical relation between the magnetic field and the cloud density, $B\propto n_{\rm{H_{2}}}^{0.5}$, for number densities up to 10$^6$~cm$^{-3}$. Later this relation was extended up to number densities of 10$^{11}$~cm$^{-3}$ (Vlemmings \cite{vle08}). Hence, by applying this relation in the case of W75N, we find that the methanol masers occur in material $\sim$50 times denser than OH, because Hutawarakorn et al. (\cite{hut02}) found a magnetic field of 7\,mG.\\
\indent The dynamical importance of the magnetic field can be quantified by considering the magnetic pressure and the dynamics pressure of the gas with velocity $v$ and density $\rho$. Comparing the two pressures we can write
\begin{equation}
 B_{\rm{crit}}=\sqrt{8\pi\rho v^{2}},
\label{bcrit}
\end{equation}
which defines the critical magnetic field strength. If $B<B_{\rm{crit}}$, the dynamics of the region is regulated by the gas motion, and if $B>B_{\rm{crit}}$, the dynamics is dominated by the magnetic field. Considering a number density $n_{\rm{H_{2}}}\lesssim10^{9}\,\rm{cm^{-3}}$, above which the masers are occasionally quenched (Cragg et al. \cite{cra05}), and velocity of $\sim$5~\kms ~from internal maser motion, we obtain a $B_{\rm{crit}}\lesssim25\,\rm{mG}$. As the magnetic field is larger than $B_{\rm{crit}}$ over the whole 2000\,AU area, it must play a dominant role in the dynamics of the W75N star-forming region.
\section{Conclusions}
We observed the 6.7\,GHz methanol masers in the massive star-forming region W75N with the EVN. We detected 10 methanol maser sources, four of which are new detections. The linear polarization indicates that the magnetic field is aligned with the large-scale molecular bipolar outflow. This is consistent with the W75N(B) twisted magnetic field picture sketched by Greaves \& Holland (\cite{gre98}). The detection of Zeeman-splitting indicates a magnetic field strength of the order of 50\,mG. We suggest VLA\,1 as the driving source of the large-scale molecular bipolar outflow of W75N(B).\\

\noindent \small{\textit{Acknowledgments.} GS was supported for this research through a salary from the International Max Planck Research School (IMPRS) for Astronomy and Astrophysics at the Universities of Bonn and Cologne. WV acknowledges support by the \textit{Deutsche Forschungsgemeinschaft} through the Emmy Noether Reseach grant VL 61/3-1. GS and WV thank Jos\'{e} M. Torrelles and Vincent Minier for kindly providing the VLA continuum image and methanol masers positions, respectively.}

\end{document}